\documentstyle[12pt,aasms4]{article}
\def\etal{{\rm et~al.\ }}
\def\sles{\lower2pt\hbox{$\buildrel {\scriptstyle <}
	\over {\scriptstyle\sim}$}}
\def\sgreat{\lower2pt\hbox{$\buildrel {\scriptstyle >}
	\over {\scriptstyle\sim}$}}

\received{................}
\accepted{................}

\begin{document}

\title{On the Nature of the Strong \\
Emission-Line Galaxies in Cluster Cl 0024+1654: \\
Are Some the Progenitors of Low Mass Spheroidals? \altaffilmark{1} }

\author{David C. Koo, Rafael Guzm\'an, Jes\'us Gallego \altaffilmark{2}} 
 
\affil{University of California Observatories / Lick Observatory, \\
Board of Studies in Astronomy and Astrophysics, \\
University of California, Santa Cruz, CA 95064}

\author{Gregory D. Wirth}

\affil{Department of Physics and Astronomy, University of Victoria\\
P. O. Box 3055, Victoria, BC, V8W 3P6 Canada} 

\author{koo@ucolick.org, rguzman@ucolick.org, jgm@astrax.fis.ucm.es,
wirth@uvastro.phys.uvic.ca}

\dates

\altaffiltext{1}{Based on observations obtained at the W. M. Keck Observatory,
which is operated jointly by the University of California and the
California Institute of Technology, and with the NASA/ESA {\it Hubble
Space Telescope} obtained at the Space Telescope Science Institute
which is operated by AURA under NASA contract NAS 5-2655.}

\altaffiltext{2}  {Current address: 
Dpto de Astrofisica Universidad Complutense 
Madrid, E-28040 Spain}

\newpage
\begin{center}
	\large\bf Abstract
\end{center}

        We present new size, line ratio,  and velocity width 
measurements for six strong
emission-line galaxies in the galaxy cluster,  
Cl 0024+1654,  at redshift $z \sim 0.4$. 
The velocity widths from Keck spectra 
are all narrow ($30\la\sigma\la 120$ km s$^{-1}$), with
three profiles showing double peaks.  
Four galaxies have low masses ($M\lesssim 10^{10}M_{\sun}$).
Whereas three galaxies were previously reported to
be possible AGNs, none exhibit
AGN-like emission line ratios or velocity widths. Two or three 
appear as very blue spirals with  the remainder more akin
to luminous H-II galaxies undergoing a strong burst of star
formation. We propose that after the burst subsides, these  galaxies 
will transform into quiescent
dwarfs, and are thus progenitors of some  cluster
spheroidals 
\footnote{We adopt 
the nomenclature suggested by Kormendy \& Bender (1994),
i.e.,  low-density, dwarf ellipsoidal galaxies like NGC 205 are called
`spheroidals' instead of `dwarf ellipticals'.} seen today.

%


\clearpage
\section{Introduction}

 Photometry in two distant clusters of galaxies 
by Butcher and Oemler (1978) revealed  a surprisingly
large population of blue galaxies (Butcher-Oemler effect).  
Follow-up spectroscopy 
by Dressler and Gunn (1982, 1983 [DG83])
and Dressler, Gunn, and Schneider (1985 [DGS]) were pioneering steps
towards confirming cluster membership and unraveling their nature from
spectral lines.  Among six blue galaxies observed in 3C 295, three appeared
to be active galactic nuclei (AGN) and three to have  
strong Balmer absorption lines and
negligible emission lines suggestive of a burst of star formation
(DG83). Among 14 blue 
members observed in  Cl 0024+16, three also appeared to be AGNs, 
but the remaining resembled 
local spirals with extended star formation, rather than
starbursts as in 3C295 (DGS). 

Since their
line-ratio criterion for AGNs was not foolproof and
their spectra were of low  spectral resolution (over 1000
km-s$^{-1}$), 
DGS suggested that linewidths would
be a useful check of their claimed AGNs.
This {\it Letter} presents such  linewidths as well as new 
line ratios and results of searches for blue compact nuclei
for all three AGN candidates in Cl 0024+1654
as well as for three other emission line galaxies.
None are confirmed to be AGNs. 
We adopt $h = 0.5$, $q_0 = 0.05$, and $\Lambda
= 0$, i.e., $\Omega_0 = 0.1$, as our cosmology.
Given these parameters, $L^*$ ($M_B \sim -21$)
corresponds to $r \sim 21.0$ and $1\arcsec$
spans 7.0 kpc at the cluster redshift of $z = 0.4$.

\section{New Observations}

Six galaxies are in our sample and all (except SDG 146) 
are among the bluest
galaxies in the cluster (see colors in Table 1).
Five are 
those with the largest [\ion{O}{2}] equivalent widths among the 14
blue cluster members studied by DGS.  The remaining
galaxy (SDG 173A) is not in
the catalogs of DGS;
Schneider, Dressler, and Gunn (1986; SDG); or Dressler and Gunn (1992).
For its identification, see Table 1
and Fig. 2 of Lavery, Pierce, and McClure (1992). 
It has been included because spectra taken 
with the Shane 3 m at Lick Observatory
identified it as a 
cluster member with very strong [\ion{O}{2}] emission (restframe
equivalent width EW(\ion{O}{2}) $\sim 94 \pm 10$~\AA).
Its emission line ratios (Table 1)
would satisfy the definition of an AGN by Dressler, Thompson, and
Shectman (1985), namely 1)  EW(\ion{O}{2}) $ > 3$~\AA ;
2) EW(\ion{O}{3}) or EW(H$\beta$) $ \approx$ EW(\ion{O}{2}) ; and
3) EW(\ion{O}{3}) $>$ EW(H$\beta$).

The key observations for all six galaxies 
were taken on UT 1994 October 3 and 4, using the
HIgh Resolution Echelle Spectrograph (HIRES; Vogt {\it et al.} 1994) at
the W. M.  Keck 10 m Telescope under $\sim 0\farcs8$ 
(FWHM) seeing. The $1\farcs15\times 14\arcsec$ slit yielded a FWHM spectral
resolution of 2 pixels or 8 km s$^{-1}$ on the Tektronix 2048$^2$ CCD.
All exposures were 1800s, except for SDG-231 which was observed twice.
Internal quartz lamp frames for flatfields and Th-Ar arc lamp frames
for wavelength calibration were taken after observing at each grating
setting.  The spectra were reduced using standard techniques described
in the IRAF/NOAO ``echelle'' package 
\footnote{IRAF is distributed by the National Optical Astronomy
Observatories, which is operated by the Association of Universities
for Research in Astronomy, Inc. (AURA) under cooperative agreement
with the National Science Foundation}. After flatfielding, cosmic rays
removal, correction for scattered light and sky-subtraction, the
central 5 pixels (i.e., 1\arcsec) were coadded to produce 1-D spectra
in velocity space.  Four echelle orders spanning the rest wavelength
region $\sim$3700--5100~\AA\ were typically used to measure 
linewidths for the strongest emission lines.
Although the velocity
profiles are complex,  we have, nevertheless, chosen
$\sigma$ (i.e., 0.425 FWHM) of a single Gaussian to characterize the
widths.  
Such complex profiles are also seen in
nearby dwarf amorphous galaxies by
Marlowe et al. (1995).

For SDG 125 and SDG 173A, we acquired additional Keck spectra with
the Low Resolution Imaging Spectrograph (LRIS; see Oke et al. 1995)
in the multislit mode.  The detector was a Tektronix 2048$^2$ CCD (24
$\micron$ or $0\farcs215$ pix$^{-1}$). We used
a 900 l mm$^{-1}$ grating blazed at 5000~\AA\ covering 5200 -
7000~\AA\ 
and  a 1200 l mm$^{-1}$ grating
blazed at 7500~\AA\ and covering 8500 - 9800~\AA. 
With a slitwidth of
$1\farcs23$, these gratings yielded 0.96~\AA\ pix$^{-1}$ and 0.66~\AA\
pix$^{-1}$ and an instrumental resolution of $\sim 4$~\AA\ FWHM and
$\sim 2.5$~\AA\ FWHM, respectively.  The data taken on the clear
nights of UT 1995 September 21 -- 23, consist of 10,200s in the blue
and 3600s in the red.  Seeing was variable but was estimated to be
about 0\farcs8 FWHM during most of the exposures.  The spectroscopic
reduction involved cosmic ray removal, wavelength calibration based on
night-sky emission lines, and background sky subtraction. 
The
slitlet for each target was treated as single, but relatively short
(8\arcsec - 15\arcsec) "long-slit''.  
The spectra for each galaxy are shown in Fig
1 (plate XXX).

For SDG146 and SDG183, lower resolution spectra
were obtained on UT 1996 September 6 with the 2.5 m Isaac Newton
Telescope at La Palma Observatory. The 235mm camera of the
Intermediate Dispersion Spectrograph (IDS, see Terlevich \&
Terlevich 1989) was used with the R600 IR grating that yielded
a dispersion of 1.7~\AA \  pix$^{-1}$ with a wavelength range of 
8137 - 9862~\AA.  The detector was a Tektronix 1024$^2$ CCD (24 $\micron$
and $0\farcs7$  pix$^{-1}$) and the slitwidth was $1\farcs9$, giving a
3.4~\AA \ FWHM resolution (2 pixels). The PA was selected to cover
both objects simultaneously.  Exposures totaling  4800s  were
obtained under good conditions with moderate seeing. The reductions
are similar to that applied to the Keck LRIS spectra.

Sizes for four galaxies are based on 
archive WFPC-2 images 
(GO 5453; PI E. Turner) that were reduced, described, and kindly
provided by Colley, Tyson, \& Turner (1996). The stacked images total
8400s for exposures in the F450W filter (blue) and 6600s in the F814W
(red).
For SDG 125 and 223, R. Lavery kindly provided 
High Resolution Camera (HRCAM; McClure
\etal 1989) 
images (2h $V$, 1h $R$)
taken on the Canada France Hawaii Telescope 
and used by Lavery et al. (1992). These 1024$^2$ SAIC CCD
images have a scale of 0\farcs13  pix$^{-1}$, a field of view 2.2' on a
side, and a PSF FWHM of $\sim$ 0\farcs45.  For both the HST and HRCAM
images (Fig. 1), fluxes  measured through  multiple circular apertures
were used to measure half-light radii in the F814W or $R$ filter. 

\section{Results}

The key
result is that the HIRES spectra 
show the strong emission lines to be very narrow ($< 60$
km-s$^{-1}$) for four galaxies and
moderately narrow ($\sim 120$ km-s$^{-1}$) for the remaining two (SDG
125 and 146).  The line profile of SDG 125 shows, however, 
two components each having $\sigma \sim $~55 km-s$^{-1}$.
SDG 173A and SDG 183 may also have such complex profiles.
The HIRES data
reveal  that none of the six galaxies have the large widths expected
for AGN's, though a weak, very broad component would be difficult to detect.  
For three AGN candidates (SDG 125, 173A, and 183), 
line ratios of H$\alpha$ and  [\ion{N}{2}]
versus [\ion{O}{3}] and $H_\beta$ 
from LRIS and IDS spectra  are {\it inconsistent}
with LINERS or Seyfert 2 galaxies (i.e., 
[{\ion{N}{2}]/H$_\alpha \times$ [\ion{O}{3}]/H$_\beta > 2$), 
but do match those of H-II galaxies (Veilleux \& Osterbrock 1987).

Besides spectra, another check for AGNs can be made by searching 
the $HST$ and HRCAM images 
for bright, blue, compact nuclei.
SDG 125 and 183 show no evidence for such  nuclear components. The
remaining
AGN candidate, SDG 223, is so close in size to the PSF of the
HRCAM images that this test is not reliable.  
By itself, this
morphological evidence against the AGN hypothesis is not very
compelling, since dust may obscure the presence of an AGN, but it
strengthens the conclusions based on the HIRES kinematic  and LRIS 
line-ratio data.

For our sample, Lavery \etal (1992) have already examined the HRCAM
images for morphology, with three noted to show the presence of a disk
(SDG 125, 146, and 231), 
so the velocity widths reported
above will be underestimates of their true values 
since these galaxies do not appear
edge-on.
SDG 173A is claimed to be a close
interacting system; four (SDG 125, 146, 173A, and 231) show 
evidence for tidal tails or disturbed morphology, and one appears
unresolved (SDG 223). The new $HST$ images  
provide improved ($\sim 2 \times$)
resolution,  with  SDG 146 appearing to have a blue ring of star formation
surrounding a central {\it red} nucleus; SDG 173A to be a very
distorted system strongly suggestive of tidal disruption;  SDG 183
to be a small face-on spiral with no bright blue
nucleus; and SDG 231 to be an elongated, narrow galaxy
with very blue colors and  high surface brightness. 

The half-light radii (Table 1) are also useful diagnostics.
SDG 125, 146, and 183 appear to be spirals of
normal size ($\sim$ 5 - 10 kpc in radius) while  SDG 173A and 223 
are only
\sles\ 3 kpc in size, which  match that of dwarf galaxies rather than
that of luminous ($L^*$) disk galaxies. SDG 231 appears to be highly elongated, so that the
half-light radius derived from circular apertures is 
larger than a radius measured from isophotal boundaries, but smaller
than the intrinsic one for an inclined disk.

\section {Discussion and Conclusions}

This project was motivated by the suggestion of DGS that three of the
cluster members with the strongest [\ion{O}{2}] emission lines in Cl
0024+1654 are AGN's.  Based on the narrow velocity widths of H$\beta$
and other emission lines as measured from HIRES, we conclude that
neither of the two ``certain Seyfert'' candidates (SDG 125 and 223),
nor the other ``probable AGN'' (SDG 183) are AGNs. Since the velocity
widths for an AGN can be as narrow as $\sigma \sim$ 125 km-s$^{-1}$
(de Robertis and Osterbrock 1986), SDG 125 ($\sigma \sim$ 110
km-s$^{-1}$) might still qualify as an AGN, particularly since no
inclination corrections were applied.  We have, however, additional
evidence that SDG 125 is not an AGN. First, the HIRES 
profiles show two, roughly equal components that is uncharacteristic of
known AGNs (D. Osterbrock, private communication). Second, we have
secured additional Keck spectra with LRIS in the far red that yield
line ratios of H$\alpha$ to  [\ion{N}{2}] that, combined with
[\ion{O}{3}] 
and $H_{\beta}$ ratios, are
consistent with star-formation (Veilleux \& Osterbrock 1987) rather
than LINERs or AGNs. Third, the entire galaxy
appears very blue, with no evidence from HRCAM images for a bright,
bluer, nuclear component. SDG 173A,
though not in the original sample of DGS, qualifies as a possible AGN
on the basis of
the DGS criterion.  Except for the additional evidence of a
complex morphology that suggests an interaction (Lavery et
al. 1992), 
the other arguments against SDG
125 being an AGN also apply here to SDG 173A.

Since none of the proposed AGNs in Cl 0024+1654 have been confirmed
with improved spectroscopy and imaging data, 
we conclude that the scenario of having more AGN
activity among distant clusters (DGS) is in serious jeopardy.  Those
AGN candidates found in other clusters (Dressler and Gunn 1992) and
without obvious broad emission lines should be re-examined before the
AGN evolution picture is accepted. 

So what is the nature of these cluster galaxies with very strong
emission lines? As already discovered in $HST$ images  
by Dressler et al. (1994), the blue galaxies
are {\it not} early-type spirals or large-bulge galaxies as originally suggested
by DGS. This result argues against 
the traditional view that the blue
galaxies seen as the Butcher-Oemler effect will become the S0's of
today (DGS). Instead, many appear to be late-type spirals and irregulars as
well as morphologically disturbed, possibly interacting or merging
systems. One attractive physical mechanism for providing such
morphologies has been suggested by Moore et al. (1996).  
Their simulations show 
that disk galaxies in clusters may become 
morphologically disturbed due to being ``harrassed'' by the
gravitational potential of the cluster and other nearby galaxies.
But a major unanswered  question is the nature of the 
descendents of these galaxies. 

Given the compelling evidence that we are observing very intense bursts of
star formation rather than AGN activity among several of the 
bluest cluster galaxies, we
suggest an alternative explanation: the very blue, luminous, cluster
galaxies with small sizes and very low velocity widths are not massive
galaxies, but {\it low-mass dwarfs} in a bursting phase (Babul and
Rees 1992). The bursts in such dwarfs would be so intense that any
remaining gas would be heated, expelled from the galaxy, and then
stripped during interactions with the ambient cluster gas, thus
destroying gas that might have the potential to fuel future star
formation (Babul and Rees 1992). For example, adopting the
energy input estimates for a starburst by Heckman et al. (1993),
SDG 223 is releasing through star
formation (Table 1) roughly 10$\times$ its binding energy. 
These galaxies are thus excellent candidates to be the
progenitors of cluster spheroidals.

Indeed, recent $HST$ and Keck HIRES observations of a sample of
compact, narrow emission line galaxies (CNELG) in the field have
revealed luminosity, size, color, line-ratio, kinematic, and
morphological characteristics (Koo et al. 1994, 1995; Guzm\'an et
al. 1996) virtually identical to that seen for several of the cluster
galaxies in the current sample.  These CNELGs have redshifts in the
range $0.1\la z\la 0.7$ and are thus seen at lookback times directly
comparable to that of Cl 0024+1654 at $z \sim 0.4$.

The connection of the CNELGs and current sample with low-mass dwarfs
is directly visible in a 
size versus velocity width diagram (Fig. 2). First note that the
masses for SDG 173A, 183, 223, and 231 
are in the range of low-mass dwarfs, i.e.,
$\lesssim 10^{10} M_\odot$, but we caution that SDG 183 may be a face-on spiral
so that its narrow linewidth may in part 
be due to its low inclination rather
than low-mass.  Second, this diagram has no luminosity, and can thus
be used to compare different galaxy types, independent of luminosity
evolution. With surface brightness and fading taken into account, SDG
231 and 223 are the best counterparts to the CNELGs (see Fig. 4 in
Koo et
al. 1995 and columns 7 and 8 in Table 1). 
Third, the likely counterparts in mass, size, and velocity
widths to both the CNELGs and the current sample are either irregulars
or spheroidals (such as NGC 205 rather than very low-luminosity dwarf
spheroidals like Carina). Whereas spheroidals have metallicities and
star formation histories that are fully consistent with a major
intense burst of star formation (Skillman and Bender 1995), 
irregulars show a more subdued history (see Guzm\'an et
al. 1996 for fuller discussion and references; see also recent evidence for
high [O/Fe] abundances in planetary nebula for spheroidals that do not
match that seen in irregulars and which suggest a briefer burst of
star formation [Richer, McCall, and Arimoto 1996]).

Thus, following arguments favoring CNELGs to be the progenitors of
local {\it field} spheroidals, we suggest that some of the blue galaxies
seen in distant clusters are the progenitors of the numerous
{\it cluster} spheroidals found today. Let us make a very rough estimate
of the expected number density of cluster spheroidals with masses
about $10^{9} - 10^{10} M_\odot$ formed through these intense bursts.
First we assume that the bursting phase lasts only a few $10^7$ years and
that the frequency of bursting dwarfs in clusters remains constant
over a several Gyr period. In this case, and making the further crude
assumption that the current spectroscopic sample of two or three is
but 10\% of the entire cluster sample of bursting dwarfs to fainter
limits, we may easily obtain a factor of 1000 for the conversion of
observed dwarfs in the burst phase to eventual descendents. This
translates our sample of two or three \footnote {For two or three
observed objects, the 95\% Bayesian confidence limits on the
underlying number densities are 0.30 and 7.95 objects (Kraft, Burrows,
and Nousek 1991)} bursting galaxies to many hundreds to several
thousand inert cluster spheroidals for Cl 0024+1654 by this time.
Such large numbers of inert spheroidals are consonant with claims that
60\% of the cluster galaxies brighter than $M_B = -14$ are spheroidals
in both Coma (Bernstein et al. 1995) and Virgo (Binggeli, Tarenghi, and
Sandage 1990).
We thus find a rough,
but consistent, link between the small numbers of bursting systems
seen at a given moment in distant clusters and the large number of
spheroidal descendents seen today.

In summary, new $HST$ images and Keck echelle spectra of the six strongest
emission line galaxies in Cl 0024+1654 show none to be bonafide
AGNs. Instead, the evidence strongly favors the sample to contain 
spirals of normal sizes but active star formation as well as  smaller,
low-mass dwarf
systems undergoing a major burst of star formation. With physical
properties similar to those found in local H-II galaxies and more
distant counterparts (CNELGs), the dwarfs are 
excellent candidates for being the progenitors of the numerous spheroidals
seen in rich clusters at the current epoch.

\vspace{1cm}

\centerline{\bf Acknowledgements}

	We thank M. Takamiya, K. Ing, R. Guhathakurta,
A. Szalay, and A. Connolly for help during the Keck observing;
A. Arag\'on-Salamanca for acquiring the INT observations; 
W. Colley, A. Tyson, and E. Turner for their generous permission
to use their processed $HST$ image of the cluster; R. Lavery,
M. Pierce, and R. McClure  for their 
kind permission to use their HRCAM images; D. Osterbrock for helpful
discussions; and the Keck staff
for their assistance at the telescope. We also thank
E. Telles for allowing us to use his data for HII galaxies prior to
publication and the referee for helpful suggestions. 
Funding for this work was provided by
NSF grants AST91-20005 and AST-8858203 and  NASA grant
AR-05802.01-94A. RG acknowledges funding from the Spanish MEC
fellowship EX93-27295297; 
JG from the Spanish MEC grants
PB89--124 and PB93--456 and a UCM del Amo Foundation Fellowship.

\clearpage

\clearpage

\title {Figure Captions}

{\sc fig}. 1. -- {a) Panel of {\it HST} images and Keck spectra for
the three AGN candidates found by DGS in Cl 0024+1654. The images of
SDG 125 in $V$ and $R$ were taken by Lavery \etal (1992) with
HRCAM. The contours are spaced at intervals of 0.5 mag in surface
brightness, beginning at 6 times the standard deviation of the local
sky value above background.  The H$\beta$ and [\ion{O}{3}] 5007
emission line profiles from the HIRES observations are shown with
normalized amplitudes and in km s$^{-1}$. The instrumental resolution is
$\sigma = 3.4$ km s$^{-1}$. The spectra below these are from two
separate grating settings of LRIS that show the strengths of various
emission lines and the lack of any broad emission component in the
Balmer lines.  For SDG 183, the $B$ (F450W) and $I$ (F814W) images
were taken with WFPC-2. The images for SDG 223 are also from HRCAM.
Dotted lines in the line profiles show the best
Gaussian curve that fits the average of profiles from the measured
emission lines.  Note the clear presence of two components in SDG 125
and the hint of another component in SDG 183. The $\sigma$ values are
based on single Gaussian fits.  b) Similar to a) for AGN candidate SDG
173A and two other strong emission line galaxies proposed to be
spirals by DGS. All images are from WFPC-2. }

{\sc fig}. 2. -- { The half-light radius ($R_e$) vs. velocity width
($\sigma$) diagram for the cluster and field
samples of strong emission line galaxies. The large shaded pentagons
are for the
cluster spiral galaxies SDG 125, 146, and 183; the open pentagons are 
for the remaining galaxies SDG
173A, 223, and 231. The crosses are for local H-II galaxies
from Telles (1995) and star symbols are for more distant CNELGs from Guzman et al. (1996).  The typical locations of
various other galaxy types are also shown, including E/S0 and
spheroidals galaxies (Bender, Burstein
\& Faber 1992), spirals (de Vaucouleurs et al. 1991 - RC3), and
irregulars (RC3). A few galaxies representative of today's evolved
systems are identified. The dashed
lines correspond to constant-masses  of $10^8$, $10^{10}$, and
$10^{12}M_{\sun}$ }.

\end{document}